\pdfoutput=1
\newif\ifcomment
\newif\ifdraft
\drafttrue
\ifdraft
\newif\ifextra
\newif\ifplainart
\ifplainart
\documentclass[11pt]{article}
\usepackage{authblk}
\usepackage[nochapters]{classicthesis} % nochapters
\usepackage[comma,square,numbers,sort&compress]{natbib}
\else
\documentclass[%
 reprint,
 amsmath,
 amssymb,
 aps,
]{revtex4-1}
\fi
\usepackage{dcolumn}% Align table columns on decimal point
\usepackage{bm}% bold math
\usepackage{color}
\usepackage{graphicx}
\usepackage{amsmath}
\usepackage{amsfonts}
\usepackage{amssymb}
\usepackage[T1]{fontenc}
\usepackage{mathrsfs}
\usepackage{calrsfs}
\usepackage{pslatex}
\usepackage[allcolors=red]{hyperref}
\usepackage{url}
\usepackage{lineno}
\usepackage[utf8]{inputenc}
\usepackage[separate-uncertainty=true,
table-align-uncertainty=true]{siunitx}
\usepackage{array,multirow,graphicx}
\usepackage{booktabs}
\newcommand{\pp}           {pp}

\newcommand{\pn}           {pn}
\newcommand{\nn}           {nn}
\renewcommand{\AA}         {AA}

\newcommand{\NN}           {NN}
\newcommand{\pPb}          {pPb}

\newcommand{\PbPb}         {PbPb}

\newcommand{\ppbar}        {$\rm p\bar p$}

\newcommand{\pT}           {\ensuremath{p_{\rm T}}}
\newcommand{\pt}           {\ensuremath{p_{\rm T}}}

\newcommand{\TAA}          {\ensuremath{T_{\rm AA}}}

\newcommand{\RAA}          {\ensuremath{R_{\rm AA}}}

\newcommand{\eg}           {e.g.}
\newcommand{\ie}           {i.e.}

\newcommand{\snn}          {\ensuremath{\sqrt{s_{\rm NN}}}}

\newcommand{\sigmapp}      {\ensuremath{\sigma_{\rm pp}}}
\newcommand{\sigmaNN}      {\ensuremath{\sigma_{\rm NN}}}
\newcommand{\sigmapPb}     {\ensuremath{\sigma_{\rm pPb}}}
\newcommand{\sigmaPbPb}    {\ensuremath{\sigma_{\rm PbPb}}}

\newcommand{\Ncoll}        {\ensuremath{N_{\rm coll}}}

\newcommand{\hrefurl}[1]   {\href{#1}{\url{#1}}}
\newcommand{\Refe}[1]      {Ref.~\cite{#1}}

\newcommand{\Tab}[1]       {Tab.~\ref{#1}}
\newcommand{\Fig}[1]       {Fig.~\ref{#1}}

\newcommand{\Eq}[1]        {Eq.~(\ref{#1})}
\newcommand{\Sec}[1]       {Sec.~\ref{#1}}

\newcommand{\Hpythia}      {{\sc HG-PYTHIA}}
\newcommand{\Hijing}       {{\sc hijing}}
\newcommand{\Pythia}       {{\sc pythia}}

\newcommand{\com}[1]       {}

\begin{document}
%\preprint{AAPM/123-QED}
%%%%%%%%%%%%%%%%%%%%%%%%%%%%%%%%%%%%%%%%
\ifplainart
\title{Centrality dependence of electroweak boson production in \PbPb\ collisions at the LHC}
%\author{Constantin Loizides, Florian Jonas}
\author[1,2]{Florian Jonas}
\author[2]{Constantin Loizides}  %\footnote{email: loizides@cern.ch}
\affil[1]{\small Westfälische Wilhelms-Universität, Institut für Kernphysik, Münster, Germany}
\affil[2]{\small ORNL, Physics Division, Oak Ridge, TN, USA}
\date{\small \today}
\else
\title{Centrality dependence of electroweak boson production \\ in \PbPb\ collisions at the LHC}
\author{Florian Jonas}
\affiliation{Westfälische Wilhelms-Universität, Institut für Kernphysik, Münster, Germany}
\affiliation{ORNL, Physics Division, Oak Ridge, TN, USA}
\author{Constantin Loizides}
\affiliation{ORNL, Physics Division, Oak Ridge, TN, USA}
\date{\today}
\fi
%%%%%%%%%%%%%%%%%%%%%%%%%%%%%%%%%%%%%%%%
\begin{abstract}
%\textbf{Abstract:} 
Recent data on the nuclear modification of W and Z boson production measured by the ATLAS collaboration in \PbPb\ collisions at $\snn=\SI{5.02}{TeV}$ show an enhancement in peripheral collisions, seemingly contradicting predictions of the Glauber model. 
The data were previously explained by arguing that the nucleon--nucleon cross section may be shadowed in nucleus--nucleus collisions, and hence suppressed compared to the proton--proton cross section at the same collision energy.
This interpretation has quite significant consequences for the understanding of heavy-ion data, in particular in the context of the Glauber model.
Instead, we provide an alternative explanation of the data by assuming that there is a mild bias present in the centrality determination of the measurement; on the size of the related systematic uncertainty.
Using this assumption, we show that the data is in agreement with theoretical calculations using nuclear parton distribution functions.
Finally, we speculate that the centrality dependence of the W$^-$/W$^{+}$ ratio may point to the relevance of a larger skin thickness of the Pb nucleus, which, if present, would result in a few percent larger \PbPb\ cross section than currently accounted for in the Glauber model and may hence be the root of the centrality bias.
\end{abstract}
%%%%%%%%%%%%%%%%%%%%%%%%%%%%%%%%%%%%%%%%
\maketitle
%%%%%%%%%%%%%%%%%%%%%%%%%%%%%%%%%%%%%%%%
\section{Introduction}
%%%%%%%%%%%%%%%%%%%%%%%%%%%%%%%%%%%%%%%%
Recent data by the ATLAS collaboration on W~\cite{Aad:2019sfe} and Z~\cite{Aad:2019lan} boson production in \PbPb\ collisions at $\snn=5.02$~TeV are significantly more precise than earlier measurements at lower centre-of-mass energy~\cite{Chatrchyan:2011ua,Chatrchyan:2012nt,Aad:2012ew,Chatrchyan:2014csa,Aad:2014bha}.
%ALICE forward: \cite{Acharya:2020puh}
As commonly done, possible modifications of yields in \PbPb\ collisions from nuclear effects were quantified with the nuclear modification factor
\begin{equation}
%\RAA^{\,\rm cent}(\pT,\eta) = 
%\frac{\dd^{2}Y^{\rm cent}_{\rm AA}/\dd\pT\dd\eta}{\Ncoll^{\rm cent}\;\dd^{2}Y_{\rm pp}/\dd\pT\dd\eta} =
%\frac{\dd^{2}Y^{\rm cent}_{\rm AA}/\dd\pT\dd\eta}{\TAA^{\rm cent}\;\dd^{2}\sigmapp/\dd\pT\dd\eta}
\RAA^{\,i,\rm cent} = 
\frac{Y^{i,\rm cent}_{\rm AA}}{\Ncoll^{\rm cent}\;Y^i_{\rm pp}} =
\frac{Y^{i,\rm cent}_{\rm AA}}{\TAA^{\rm cent}\;\sigmapp^i}
\label{eq:RAA}
\end{equation}
by comparing the measured yields of bosons of type $i$=W$^+$, W$^-$ or Z in \PbPb\ collisions to the yields measured in inelastic \pp\ collisions scaled by the number of incoherent nucleon--nucleon~(\NN) collisions for a given collision centrality.
The determination of collision centrality %from the experimental event selection based on multiplicity or transverse energy 
as well as the calculation of the number of collisions $\Ncoll$ and the nuclear overlap $\TAA$ relies on the Glauber model~\cite{Miller:2007ri,dEnterria:2020dwq}.
In the Glauber model a nucleus--nucleus~(\AA) collision is approximated in the eikonal formalism, where nucleons in the projectile travel along straight lines and undergo multiple independent collisions with nucleons in the target using the inelastic \NN\ cross section $\sigmaNN$\com{ at a given centre-of-mass energy} as inter-nucleon interaction strength, so that $\Ncoll=\TAA\,\sigmaNN$. 
The centrality, which is usually given as a percentage of the total nuclear interaction cross section, is commonly determined by fitting the expectations from the Glauber model coupled with simple mechanisms for particle production to measured multiplicity or transverse energy distributions.
The absolute scale of the centrality is determined by an anchor point (AP), which relates \eg\ a measured multiplicity to a specific centrality. 
%The total number of collisions $\Ncoll$ depends on the nuclear overlap $\TAA$ between the two colliding nuclei as $\TAA\,\sigmaNN$.

The data on the modification factors for W and Z bosons were found~\cite{Aad:2019sfe,Aad:2019lan} to exceed state-of-the-art perturbative calculations at next-to-leading~(NLO) order using nuclear-modified parton distribution functions~(PDFs) by $1$--$3$ standard deviations, in particular for peripheral collisions. 
%Eskola et al.\ showed~\cite{Eskola:2020lee} that data and calculations can be brought to agreement using a fitted value of $\sigmaNN=41.5^{+16.2}_{-12.0}$~mb instead of $70\pm5$~mb, (or instead of the more precise value of $67.6\pm0.6$~mb~\cite{Loizides:2017ack}) %or the latest recommended value of $67.3\pm1.2$~mb~\cite{dEnterria:2020dwq} 
%and that such a suppressed cross section is consistent with the expectations from an eikonal minijet model incorporating nuclear shadowing~(see \Fig{fig:shcross}).
%The authors hence question the standard approach of using the measured inelastic \pp\ cross section as input to Glauber calculations, since it may lead to a misinterpretation of the experimental data.
Eskola et al.\ argued~\cite{Eskola:2020lee} that the data can be explained using a fitted value for $\sigmaNN$ that is significantly smaller than the reference value used at $\snn=5.02$~TeV.
Since the extracted value was found to be consistent with the expectations from an eikonal minijet model incorporating nuclear shadowing, their findings question the standard approach of using the measured inelastic \pp\ cross section as input to Glauber calculations. 

In this paper we provide an alternative explanation for the data by assuming a bias of the anchor point used for the centrality determination, where the size of the bias is compatible with the related systematic uncertainties of the measurement.
We furthermore explore the influence of the neutron skin on the centrality dependence of $W^+$ and $W^-$ bosons, as well as potential consequences for the overall PbPb cross section used in the Glauber model. 
%The sensitivity of $W$ boson production to the presence of a neutron skin was also shown in \Refe{Paukkunen:2015bwa} using an optical Glauber model.  Furthermore, the author suggests that in the future, such measurements could be used to benchmark different centrality definitions at the LHC.
The paper is structured as follows: 
In \Sec{sec:shadowing} we first discuss in more detail the nuclear shadowing explanation offered by Eskola et al.~\cite{Eskola:2020lee} and its potential shortcomings. % maybe a bit strong
In \Sec{sec:bias}, we introduce the assumed AP bias, and show that it counteracts the multiplicity bias usually present for high-$\pT$ probes.
In \Sec{sec:model}, we construct a reference model for the $\RAA$ that includes no nuclear effects by incorporating the centrality dependence of the isospin and the deduced residual bias, and compare the reference model with the data.
In \Sec{sec:neutronskin}, we discuss the influence of the neutron skin on the measurement.
Finally, we summarize our findings in \Sec{sec:summary}.

\begin{figure}[t!]
\begin{center}
\includegraphics[width=\linewidth]{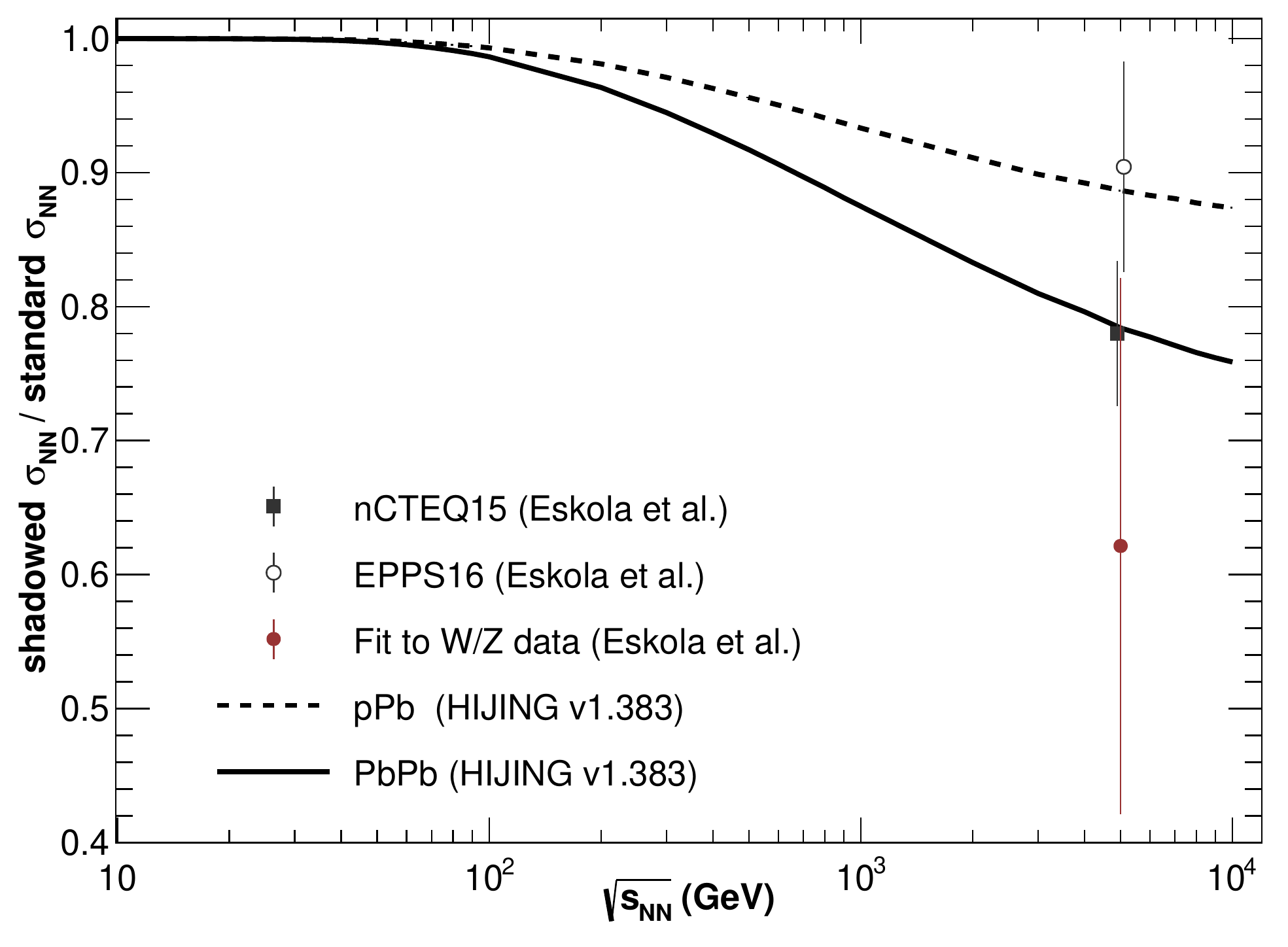}
\caption{Ratio of shadowed over standard inelastic cross section in \pPb\ and \PbPb\ collisions as function of $\snn$ calculated with \Hijing~\cite{Gyulassy:1994ew}. %v1.383.
The results from Eskola et al.~\cite{Eskola:2020lee}, \ie\ the fit to the boson data as well as the computed cross sections at 5.02~TeV with their minijet model, are also shown. The reported values were transformed from asymmetric to symmetric uncertainties, and then combined. For the calculations, additionally it was assumed that the results from different factorization scales were independent of each other. 
} %macro from sw/biasraa/p8/plotCross.C 
\label{fig:shcross}
\end{center}
\end{figure}

%%%%%%%%%%%%%%%%%%%%%%%%%%%%%%%%%%%%%%%%
\section{Shadowing of the inelastic nucleon--nucleon cross section?}
\label{sec:shadowing}
%%%%%%%%%%%%%%%%%%%%%%%%%%%%%%%%%%%%%%%%
As mentioned above, Eskola et al.\ showed~\cite{Eskola:2020lee} that data and calculations on the boson $\RAA$ can be brought to agreement, when using a fitted value of $\sigmaNN=41.5^{+16.2}_{-12.0}$~mb instead of $70\pm5$~mb (or instead of the more precise value of \mbox{$67.6\pm0.6$~mb~\cite{Loizides:2017ack}}) \com{or the latest recommended value of \mbox{$67.3\pm1.2$~mb~\cite{dEnterria:2020dwq}}}.
The extracted cross section turned out to be consistent with the expectations from an eikonal minijet model incorporating nuclear shadowing using the EPPS16~\cite{Eskola:2016oht} or nCTEQ15~\cite{Kovarik:2015cma} nuclear PDFs~(see \Fig{fig:shcross}).
Using a different value for $\sigmaNN$ than the measured inelastic \pp\ cross section as input to Glauber calculations would break binary-collision scaling, and hence question the common approach to construct a reference for the number of hard collisions obtained in \AA\ collisions using the Glauber model.
Furthermore, the potential shadowing of the inelastic cross section is collision energy and system dependent. %, and hence would complicate data-and-model comparisons. 
As demonstrated in \Fig{fig:shcross}, which also shows the ratio of shadowed over standard inelastic cross section in \pPb\ and \PbPb\ collisions as function of $\snn$ calculated with the \Hijing~\cite{Gyulassy:1994ew}\com{v1.383} minijet model,
the effect increases with increasing collision energy, and is roughly twice as strong for \PbPb\ than for \pPb\ collisions.
For \PbPb\ collisions at 5.02~TeV, \Hijing\ predicts a similar suppressed cross section as the nCTEQ15 calculation, while the EPPS16 calculation expects a smaller suppression, both consistent with the suppressed results but with no suppression. 
At the highest RHIC energy of $\snn=0.2$~TeV, the expected suppression is already below 5\%.

The central value of about $41.5$~mb for the reduced cross section is rather small, only about $60$\% of the typical $\sigmaNN$ used at 5.02~TeV, and of similar magnitude as the unshadowed cross section at the highest RHIC energy.
Using $\sigmaNN=41.5$~mb essentially squeezes \AA\ collisions into a smaller range of impact parameters, and relative change of $\TAA$ to the $\TAA$ computed at 67.3~mb increases with decreasing centrality reaching factor $\sim 1.5$ for most peripheral collisions.
This would result in an observable change for the nuclear modification factors measured previously at 5.02~TeV, which primarily have been measured for hadrons and jets, would be smaller, by up to 35\% in most peripheral collisions.

For nominal $\sigmaNN$ derived from inelastic pp collision data~\cite{Loizides:2017ack}, Glauber MC calculations give $\sigmaPbPb^{\rm MC}=7.55\pm0.15$~b at $\snn=2.76$\,TeV and $\sigmapPb^{\rm MC}=2.08\pm0.03$~b at $\snn=5.02$\,TeV, in good agreement with the measured values of $\sigmaPbPb=7.7\pm0.6$~b~\cite{ALICE:2012aa}, and $\sigmapPb=2.08\pm0.08$~b~\cite{Abelev:2014epa,Khachatryan:2015zaa}, respectively. %7.57 was found with Pbpnrw/Pbpnrw
%At 5.02~TeV, $\sigmaPbPb=7.62\pm0.15$~b for $\sigmaNN=67.3\pm1.2$~mb. %For 41mb, it is 7.24 % for $\snn=61.2]\pm1.2$ at 2.76 
Reducing the input $\sigmaNN$ by 40\% for \PbPb\ and 20\% for \pPb\ collisions, reduces the respective total cross sections computed by Glauber MC by about 5\%, which is about the relative size of the respective systematic uncertainties. %of 7.2mb for \PbPb\ and 1.98 for \pPb\ collisions. 
%The quoted uncertainties account for the propagated $\sigmaNN$ uncertainties plus, in quadrature, the resulting uncertainty of independently varying the density parameters by 1~standard deviation, which amount to about 2\% and 1.5\% for \PbPb\ and \pPb\ collisions, respectively, and are dominated by the uncertainty of the neutron skin width.
At high collisions energies~(and with large nuclei) the measurement of the total \AA\ cross section is complicated by the huge background generated by the electromagnetic fields of the incoming nuclei, making a direct measurement of this effect difficult.

%%%%%%%%%%%%%%%%%%%%%%%%%%%%%%%%%%%%%%%%
\section{Multiplicity and anchor point bias}
\label{sec:bias}
%%%%%%%%%%%%%%%%%%%%%%%%%%%%%%%%%%%%%%%%
As discussed above, an explanation of the data through the assumption of significant shadowing the inelastic \NN\ cross section $\sigmaNN$ implies rather strong consequences for interpreting the data using the Glauber model and would affect earlier measurements in \PbPb\ collisions at $\snn=5.02$~TeV, in particular precise measurements of charged particle $\RAA$ in peripheral collisions~\cite{Khachatryan:2016odn,ATLAS:2017rmz,Acharya:2018njl}. 
It is therefore important to explore alternative explanations of the data, focusing on the measurement itself and possible biases rather than to conclude immediately assuming physical effects. %, which we want to do in this paper. 
Before discussing the possible biases and their effect on the data, we mention a few details of the measurement and the associated systematic uncertainties.
We focus on the Z-boson measurement~\cite{Aad:2019lan}\com{ of the ATLAS collaboration}, which is reported for minimum-bias collisions, \ie\ integrated over the 0--100\% centrality range. 
Even though many arguments are also valid for the W-boson measurement~\cite{Aad:2019sfe}, we choose the $Z$-boson measurement as a starting point due to the fact that isospin effects are small~(see \Tab{tab:values}).

Even a centrality-integrated measurement depends on the estimate of the total hadronic sample, and hence implicitly on the centrality determination~(unless a cross section is directly measured in \PbPb\ collisions).
The Z-boson yield
$ %\begin{equation}
Y = \frac{N\,\varepsilon}{n_{\rm evts}\,c}
$ %\end{equation}
was obtained from the measured raw Z-boson candidate yield $N$, corrected for signal impurity and inefficiency $\varepsilon$, divided by the number of hadronic \PbPb\ events $n_{\rm evts}$, corrected for trigger and event selection inefficiencies $c$.
In particular, for more peripheral collisions, various effects influence the measured yield, such as the determination of the total event sample, contributions from~(out-of-time) pileup collisions, as well as contributions from electromagnetic background sources.
The determination of the total event sample was done anchoring the 0--80\% centrality class with a precision about 1.4\%, and then extrapolated to the 80--100\% class using the Glauber model. % and accounting for the background sources.
Except for the contribution from pileup, which was quantified to be less than 2\% in the most peripheral~(80--100\%) class, the background effects were accounted for, and uncertainties reaching up 7\%~(stat.) and 8\%~(sys.) were assigned. 
The uncertainty of the trigger and event selection were quantified to reach values of up to 5\%. 

From the above discussion, it is clear that the overall normalization of the \PbPb\ collision data is crucial.
An effective precision of better than a few percent in the total number of hadronic collisions needs to be achieved.
Underestimating the number of hadronic collisions, will reduce the effective total cross section $\sigmaPbPb$~(and $\TAA$) by similar amount, and may be mistakenly argued to result from a shadowed cross section. 
Since the total event sample is determined using an anchor point in multiplicity or transverse energy distributions~(at \SI{80}{\percent} in the transverse energy distribution measured at $3.1<|\eta|< 4.9$ for the ATLAS measurements), we refer to a bias in the determination of the total event sample as {\it anchor point}~(AP) bias in the following. 
In addition to a potential bias on the estimated total sample, a bias due to the ordering of events in multiplicity classes can arise. 
This bias is known as {\it multiplicity bias} and discussed in detail in \Refe{Morsch:2017brb}.
We now explain the effects of the two biases and their interplay, first briefly outlining the multiplicity bias and its effect on the Z-boson yield and then moving on to the effects of the assumed AP bias. 

\begin{figure}[t!]
\begin{center}
\includegraphics[width=\linewidth]{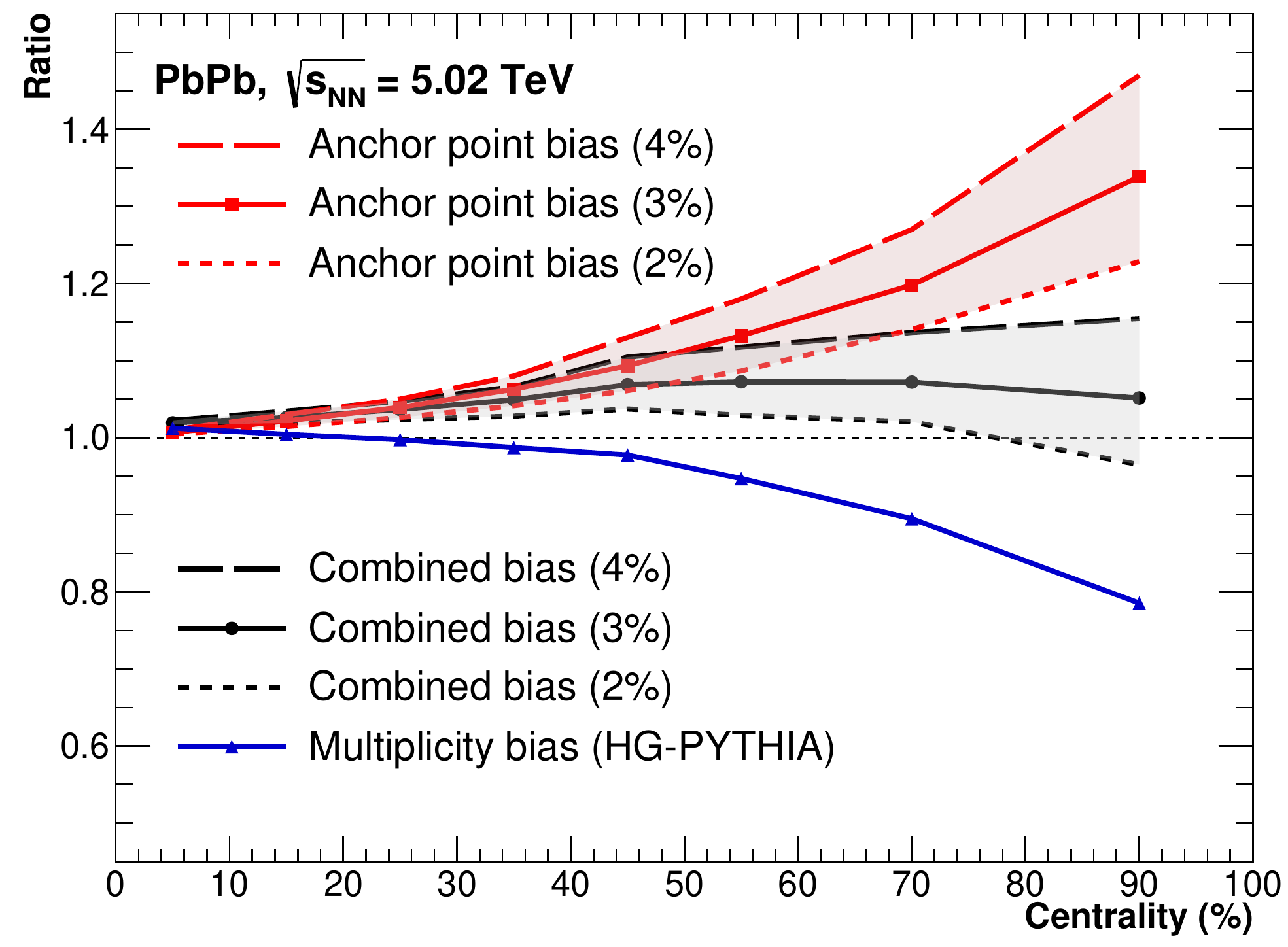} 
\caption{Illustration of the combined bias~(black) resulting from an AP bias of 2--4\% and the multiplicity bias.
The effect of the AP bias~(red) is expressed as the ratio of $\TAA$ computed with and without the bias for \PbPb\ collisions at 5.02~TeV using {\sc TGlauberMC}~\cite{Loizides:2017ack} with $\sigmaNN=\SI{67.6}{mb}$. The multiplicity bias~(blue) from the centrality determination is estimated from \mbox{\Hpythia~\cite{Morsch:2017brb}}. 
} %macro from macro/plotbias.C 
\label{fig:bias}
\end{center}
\end{figure}

To determine the centrality, events are typically ordered according to multiplicity or transverse energy measured in certain rapidity intervals~(\eg\ see~\Refe{dEnterria:2020dwq}).
%The classification can lead to event samples for which the properties of the underlying binary \NN\ collisions deviates from those of unbiased \pp\ collisions~\cite{Morsch:2017brb}. 
%The number of hard processes is suppressed for increasingly peripheral \AA\ collisions because of a simple geometrical bias: the probability for collisions increases proportional to $b$ while the nuclear density decreases, leading to an increased probability for more-peripheral-than-average \NN\ collisions.
The centrality classification, which relies on measurements dominated by soft particle production, biases the average multiplicity of individual \NN\ collisions, and hence can affect the normalization of yields of collisions dominated by hard processes due to a correlation between soft-and-hard particle production~\cite{Morsch:2017brb}. 
Hard scatterings are more probable in central \NN\ collisions with large partonic overlap thereby leading a large-underlying event activity,\com{ and since hard processes are dominated by the production of jets that fragment into a large number of final hadrons} so that a peripheral \PbPb\ event with a hard scattering often has a hadronic activity much larger than the average in its centrality class.
%Since the \AA~(and \pA) centrality determination is based on ordering the measured multiplicity or summed energy in the event, 
Peripheral nuclear events with a hard scattering can thereby be wrongly assigned to a more central class, leading to a seemingly suppressed quantity in peripheral centrality classes.
The correlation between hard scatterings and the underlying event, which first was observed for jet production\com{ by the UA1 collaboration} in \ppbar\ collisions at $\sqrt{s}=540$~GeV~\cite{Arnison:1983gw}, is similarly present for Z boson production~(\eg\ see Fig.\ 24 of \Refe{Aad:2014jgf}), and hence should appear also in peripheral \PbPb\ collisions~(in form of a decreasing \RAA with decreasing centrality). 
The magnitude of the multiplicity bias on the Z yield is shown in \Fig{fig:bias} resulting in a reduced yield by up to 20\% in the most peripheral centrality class.
It was computed using the bias factor from the \Hpythia\ model~(Fig.~3 of \Refe{Morsch:2017brb}) taking into account that $\Ncoll$ determined from so-called Glauber fits of the data already corrected for part of the bias~(relevant only for the most peripheral class, see Fig.~1 of \Refe{Acharya:2018njl}).
The \Hpythia\ model, which uses the \mbox{\Hijing~\cite{Gyulassy:1994ew}} multi-parton model to determine the number of hard \NN\ collisions in a nuclear collision and the \Pythia~\cite{Sjostrand:2006za} event generator to generate the corresponding \NN\ events, was previously used to explain the apparent suppression of the nuclear modification factor of high-$\pt$ particle production in peripheral collisions~\cite{Acharya:2018njl}.

To study the sensitivity of the measurement to the precision of the AP determination, we perform a Glauber simulation for \PbPb\ collisions at $\snn=5.02$ TeV (using {\sc TGlauberMC}~\cite{Loizides:2017ack} with $\sigmaNN=67.6$~mb).
In determining a biased value for $\TAA=\Ncoll/\sigmaNN$, it is assumed that the AP, nominally at 100\% in our calculation, is shifted by a few percent with respect to the true value, \ie\ pretending that a few percent of the most peripheral events are missed without correcting for it.
%In our study, we set the AP at 100\% to compute the nominal $\TAA$ values in intervals of centrality.
The ratio of $\TAA$ obtained by missing 2--4\% of the most peripheral events to the nominal $\TAA$ in intervals of centrality is shown in \Fig{fig:bias}.
The effect on $\TAA$~(or rather $\Ncoll$) is quite large: 
Already in the 50--60\% class a 10\% bias on the yield can be expected, which enlarges to 30\% for the most peripheral~(80--100\%) class.

\begin{table*}[t!] 
%\begin{center}
\caption{Calculated fiducial cross sections $\sigma_{\rm \pp}^{\rm fid}$ for W and Z bosons using the MCFM~\cite{Campbell:2019dru} program at NLO with the NNPDF3.1~\cite{Ball:2017nwa} for \pp, \pn\ and \nn\ collisions at $\sqrt{s} = \SI{5.02}{TeV}$, as described in the text.
The average value of $\RAA$ due to the isospin dependence $R_{\rm iso}$ was obtained from these cross sections using the fractions $Z^2/A^2=0.155$, $(A-Z)^2/A^2=0.478$ and $2Z\,(A-Z)/A^2=0.367$ as weights.
The systematic uncertainties on the cross sections are about 1\% or lower, and cancel in $R_{\rm iso}$. 
The average value of $\RAA$ including nuclear effects, $\RAA^{\rm th}$, was obtained by taking the ratio of the cross section calculated with nuclear PDFs, nNNPDF2.0~\cite{AbdulKhalek:2020yuc} and EPPS16~\cite{Eskola:2016oht} relative to the respective proton PDFSs, NNPDF3.1~\cite{Ball:2017nwa} and CT14~\cite{Dulat:2015mca}. 
The uncertainties of $\RAA^{\rm th}$ are obtained from the respective PDF uncertainties propagated as uncorrelated to the ratio.}
\label{tab:values}
\begin{ruledtabular}
\begin{tabular}{l||c|c|c|c||cc} 
       & $\sigma_{\rm \pp}^{\rm fid}$ (pb) & $\sigma_{\rm \pn}^{\rm fid}$ (pb) &  $\sigma_{\rm \nn}^{\rm fid}$ (pb) & $R_{\rm iso}$ & $\RAA^{\rm th,nNNPDF2.0}$ & $\RAA^{\rm th, EPPS16}$\\ \hline
 W$^+$ & 2233  & 1889  & 1554  & 0.81 & $0.725\pm0.022$ & $0.753\pm 0.012$\\
 W$^-$ & 1382  & 1614  & 1855  & 1.20 & $1.085\pm0.035$ & $1.110\pm0.017$\\
 Z     & 357.7 & 361.9 & 364.5 & 1.01 & $0.933\pm0.030$ & $0.960\pm 0.012$\\
 \end{tabular}
 \end{ruledtabular}
%\end{center}
\end{table*}

%The effect of the AP bias can hence be expected to be counteracted by the effect of the multiplicity bias.
The combined bias~($c_{\rm bias}$) on the expected Z boson nuclear modification factor, which can be estimated as the product of the anchor and multiplicity biases, is shown in \Fig{fig:bias} for an assumed AP bias of 2--4\%.
%For an AP bias of 3\%, the residual effect on the nuclear modification factor is hence a rather small, almost constant, enhancement of a few percent for peripheral collisions.
Already an AP bias of 2\% can essentially counter-act the multiplicity bias, while a 3\% or 4\% bias on the anchor point
leads to a small, gradually increasing enhancement of up to about 5\% and 10\% for peripheral collisions.
Hence, an alternative explanation of the enhancement seen in the data that does not rely on the shadowing of the inelastic cross section may simply be that about 3\% of the hadronic events were not accounted for in the ATLAS measurement.
In the rapidity-dependent centrality-integrated results, these missing events would reflect as a normalization issue, while in the centrality dependent measurements, they would lead to an approximate constant, depending on the actual size of the AP bias also slightly rising, enhancement with decreasing centrality.

During the writing of this paper, a new Z-boson measurement of the CMS collaboration~\cite{Sirunyan:2021vjz}, which uses about 3 times more \PbPb\ collision data than~\Refe{Aad:2019sfe} was published, which exhibits a slightly falling $\RAA$ with decreasing centrality, as expected by \Hpythia\ and in contrast to the findings by ATLAS. %, rather than slightly rising with decreasing centrality .
To quantitatively check the consistency between the ATLAS and CMS measurements, we account for the fiducial acceptance imposed by the lepton daughter selection of $|\eta^l|<2.5$ for the ATLAS measurement, which in \Pythia\ amounts to 72\%. 
%ATLAS ptl>20 etal<2.5 66<mass<116
%CMS pt>20, etal<2.1 (electron matters) but corrected to full acceptance
Integrating the data from Fig.~4 of \Refe{Aad:2019sfe} and comparing to the integrated data from Fig.~2~(or the 90\% point from Fig.~4) of~\Refe{Sirunyan:2021vjz} converted into yields per minimum-bias \PbPb\ collision, we find an approximately 5\% higher Z-boson yield measured by ATLAS compared to that measured by the CMS collaboration; corroborating our point.

While we focused on explaining the effects of the two biases on the Z-boson measurement, they similarly affect the W-boson measurement, in addition to the contribution of the isospin and its centrality dependence, which is stronger for W$^{\pm}$, as discussed below.

%%%%%%%%%%%%%%%%%%%%%%%%%%%%%%%%%%%%%%%%
\section{Reference model for \RAA in absence of nuclear effects}
\label{sec:model}
%%%%%%%%%%%%%%%%%%%%%%%%%%%%%%%%%%%%%%%%
\begin{figure}[th!]
\begin{center}
\includegraphics[width=\linewidth]{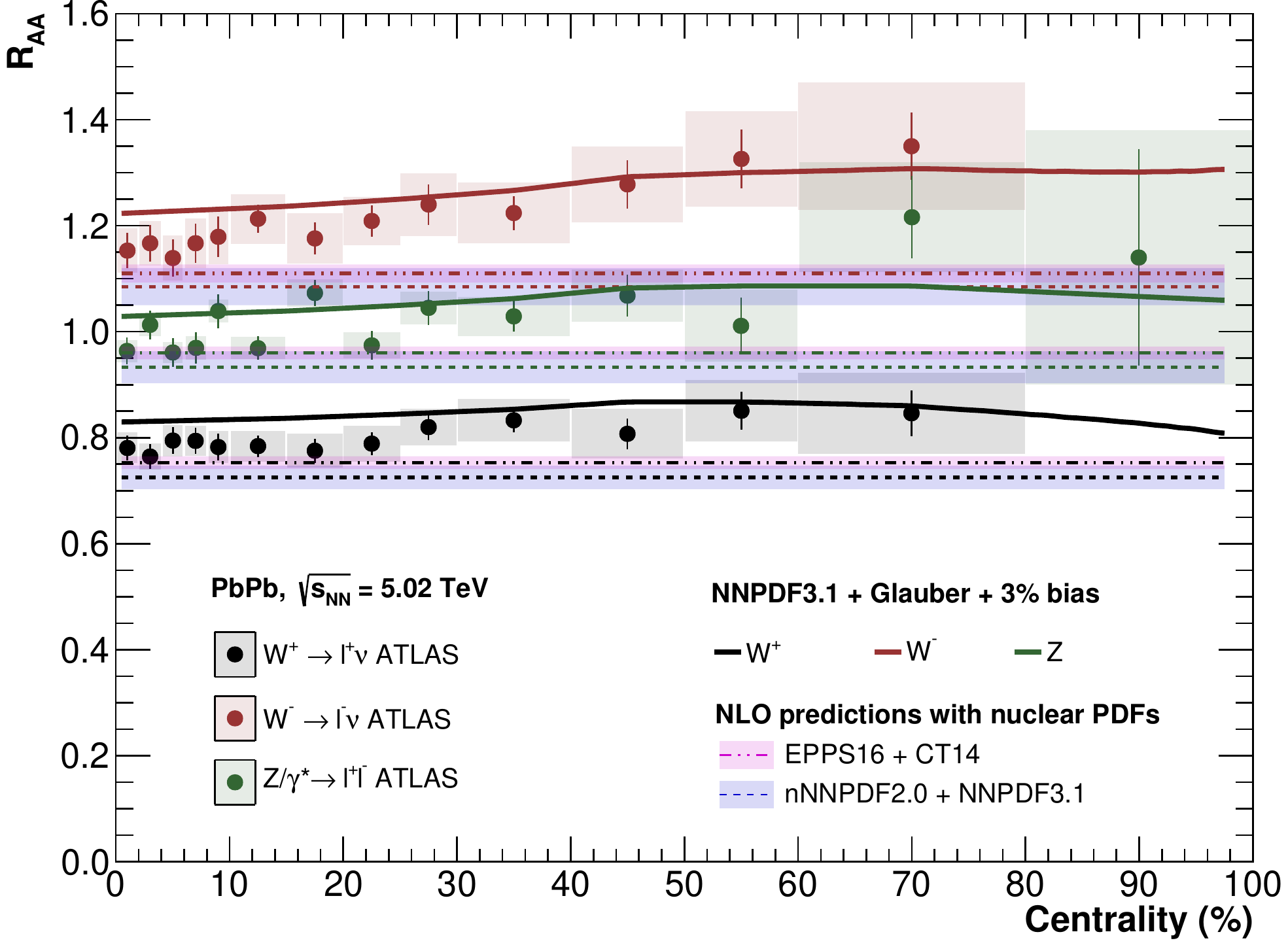}
\caption{Measured W$^+$, W$^-$ or Z boson \RAA~\cite{Aad:2019sfe,Aad:2019lan} as a function of centrality compared to various calculations. 
The solid lines show the expected value for $\RAA$ without nuclear modification, $\RAA^{\rm ref}$ from \Eq{eq:model}, using the NNPDF3.1~\cite{Ball:2017nwa} a NLO cross section calculations with the centrality dependent weights from Glauber to describe the isospin dependence and an anchor bias of 3\%, as explained in the text.
The dashed lines with the shaded band denote the $\RAA^{\rm th}$ obtained from NLO calculations incorporating nuclear effects via the use of the nNNPDF2.0~\cite{AbdulKhalek:2020yuc} and EPPS16~\cite{Eskola:2016oht} PDFs given in \Tab{tab:values}.}
\label{fig:raa}
\end{center}
\end{figure}
Using the bias factor $c_{\rm bias}$, we can compute the reference of the nuclear modification factor in absence of nuclear effects for bosons of type $i=$W$^+$, W$^-$ or Z as
\begin{equation}
\RAA^{\,i,\rm ref}(C) = \RAA^{i,\rm{iso}}(C) \, c_{\rm bias}(C) \,,
\label{eq:model}
\end{equation}
 where 
\begin{equation}
\RAA^{i,{\rm iso}}(C)= \left(f_{\rm \pp}(C)\sigma^{i,{\rm fid}}_{\rm \pp} + f_{\rm \pn}(C)\sigma^{i{,\rm fid}}_{\rm \pn} + f_{\rm \nn}(C)\sigma^{i,{\rm fid}}_{\rm \nn}\right)/\sigma^{i,{\rm fid}}_{\rm \pp}
\label{eq:model2}
\end{equation}
describes the iso\-spin dependence of $\RAA$ versus centrality~($C$) in the absence of nuclear effects. %, and $\alpha^i_{\rm mod}$ is a normalization factor accounting for possible nuclear modification.
%In absence of nuclear effects, $\RAA^{\,i,\rm ref.}(C)=\RAA^{i}$.
The boson production cross sections $\sigma^{i,{\rm fid}}_{\rm pp}$, $\sigma^{i,{\rm fid}}_{\rm pn}$ and $\sigma^{i,{\rm fid}}_{\rm nn}$ in \pp, \pn\ and \nn\ collisions at $5.02$~TeV, given in \Tab{tab:values}, were calculated at next-to-leading order (NLO) using the MCFM~\cite{Campbell:2019dru} program and NNPDF3.1~\cite{Ball:2017nwa} parton distribution functions.
The same fiducial selections as for the data were applied: $\pT^{l,\nu}>25$~GeV/$c$, $\eta^l<2.5$ and $m_{\rm T}>40$~GeV/$c^2$ for the W, and $\pt^l>20$ GeV/$c$, $\eta^l< 2.5$ and $66 < m_{\rm inv} < 116$ GeV/$c^2$ for the Z bosons.
The $\sigma^{i,{\rm fid}}_{\rm pp}$ were found to describe the data in \pp\ collisions to within their respective systematic uncertainties about 1\% and lower~\cite{Aaboud:2018nic}.
In order to calculate the cross sections for \pn\ and \nn\ collisions, a neutron PDF is needed, which is obtained from the proton PDF by exploiting isospin symmetry, \ie\ switching the contributions from up and down flavors.
The differences of $\sigma^{i,{\rm fid}}$ seen for $W^{+}$ and $W^{-}$ are caused by their differing weak isospin $T_3^{W^{\pm}}=\pm 1$, where \eg\ the production of a $W^{+}$ is favored in the case of an incoming proton ($T_3^{uud}=+1/2$).
%For \pn\ collisions, the result is approximately given by the average of the respective \pp\ and \nn\ cross sections, as expected.
The fraction of \pp, \pn\ and \nn\ collisions relative to all \NN\ collisions at a given centrality was calculated using {\sc TGlauberMC}~\cite{Loizides:2017ack} with standard settings for \PbPb\ collisions at $\snn=5.02$ TeV~(\ie\ $\sigmaNN=67.6$~mb), as can be seen in Fig.~14 of~\Refe{Loizides:2017ack}.
In central collisions, the fractions are about $f_{\rm \pp}=0.16$, $f_{\rm \pn}=0.48$ and $f_{\rm \nn}=0.36$, very close to the average values of $Z^2/A^2=0.155$, $(A-Z)^2/A^2=0.478$ and $2Z\,(A-Z)/A^2=0.367$.
Due to the so-called neutron skin of Pb, \ie\ the fact neutrons dominantly populate the outer regions of the Pb nucleus, the number of collisions involving neutrons rise with decreasing centrality, resulting in fractions of about 0.05, 0.45 and 0.50, respectively, in most peripheral collisions.
Compared to the W bosons, Z bosons are expected to have a negligible isospin dependence on centrality, because the respective \pp, \pn\ and \nn\ cross sections are numerically very similar~(see \Tab{tab:values}).
A more detailed discussion of the neutron skin and the sensitivity of the data to its thickness is deferred to \Sec{sec:neutronskin}.
In central collisions, where there is no effect from the multiplicity and AP biases~(see \Fig{fig:bias}), the value of $\RAA^{\rm ref}$ is close to that of the average value of $\RAA^{\rm iso}$, which was computed using the fractions for the different collision types and the respective cross sections in \Tab{tab:values}.
%Fit 0-10 W+ = 0.782609 +/- 0.0108709
%Fit 0-10 W- = 1.1601 +/- 0.0158632
%Fit 0-10 Z = 0.986291 +/- 0.0125301
%Theory rat: 0.977477 0.962815 0.971875
%Dat/iso rat: 0.965157 0.963618 0.977126
%Dat/EPPS16 rat: 0.965157 0.963618 0.977126

\Fig{fig:raa} shows the data of the W$^+$, W$^-$ and Z boson \RAA\ measured by the ATLAS collaboration~\cite{Aad:2019sfe,Aad:2019lan} as a function of centrality.
The data are compared to our model of the expected value for $\RAA$ without nuclear modification, $\RAA^{i,{\rm ref}}$ from \Eq{eq:model}, assuming the presence of a common AP bias of 3\% in the data.
As demonstrated in the figure, the reference calculation well describes the observed upward trend of the data and is consistent with the data in peripheral collisions.
In central collisions, the data exhibit a trend to be a 3--4\% below the reference values.
However, when comparing to theoretical predictions $\RAA^{i,{\rm th}}$ that do incorporate nuclear effects (dotted lines), one finds that data and predictions are consistent within uncertainties, even though the central values of the data tend to be 3--6\% above the predictions. 
The theoretical predictions of $\RAA^{i,{\rm th}}$ are given in \Tab{tab:values} and were obtained by taking the ratio of the cross section calculated with the nNNPDF2.0~\cite{AbdulKhalek:2020yuc} and EPPS16~\cite{Eskola:2016oht} nuclear PDFs relative to the respective proton PDFs, NNPDF3.1~\cite{Ball:2017nwa} and CT14~\cite{Dulat:2015mca}.
The shown uncertainties are the 1-$\sigma$ uncertainties of the respective proton and nuclear PDFs, which were treated as uncorrelated in the ratio, while the scale uncertainties of the NLO calculations were canceled in the ratio.
%The calculations and the data are mutually consistent within uncertainties.

\begin{figure}[t!]
\begin{center}
\includegraphics[width=\linewidth]{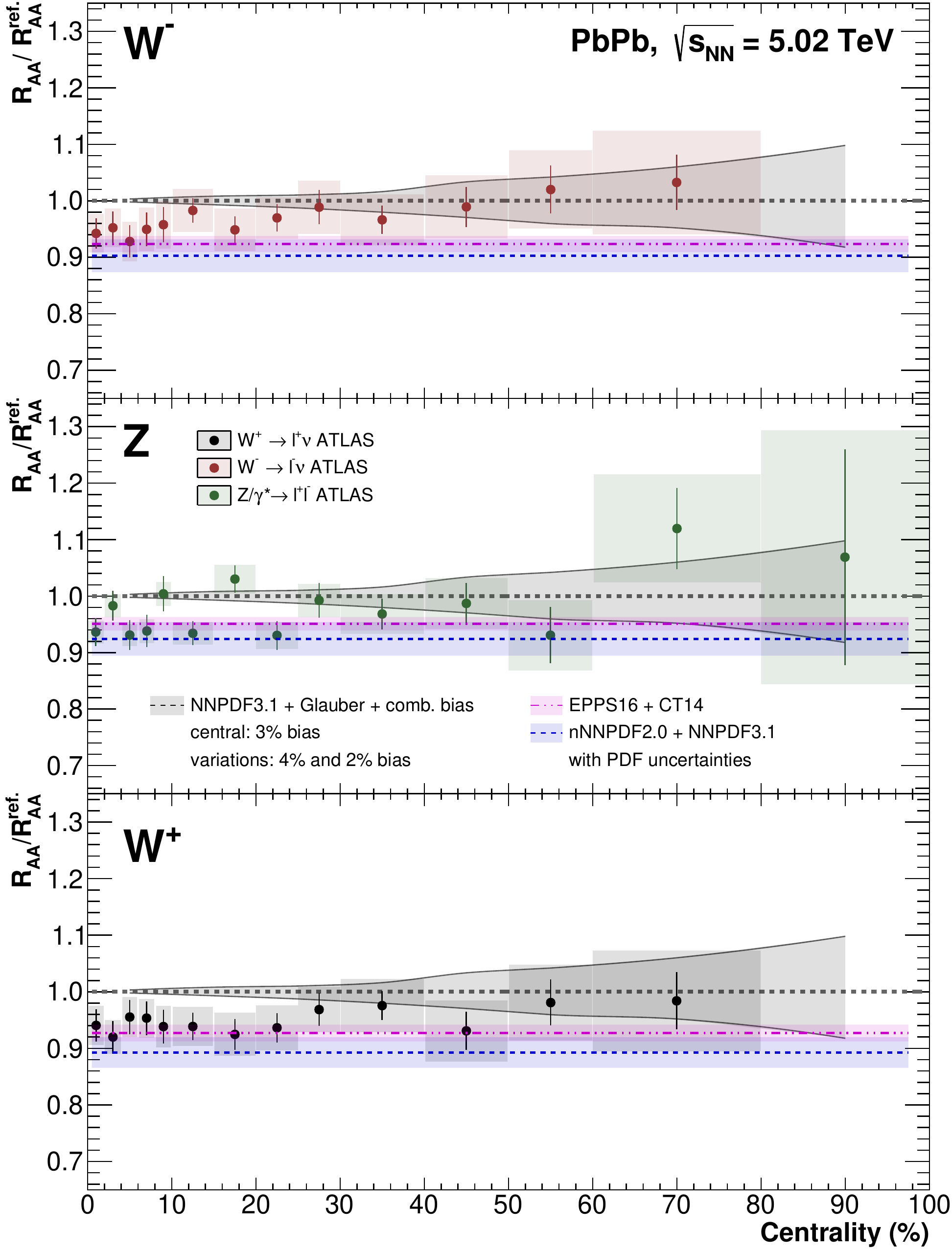}
\caption{Measured W$^+$, W$^-$ or Z boson \RAA~\cite{Aad:2019sfe,Aad:2019lan} normalized by $\RAA^{i,{\rm ref}}$, from \Eq{eq:model}, as a function of centrality.
The central value was obtained assuming an AP bias of 3\%
The band around unity denotes the change of the ratio if 2\%~(upper value) or 4\%~(lower value) had been assumed instead.
The dashed lines with the shaded band denote the $\RAA^{i,{\rm th}}$ using the nNNPDF2.0~\cite{AbdulKhalek:2020yuc} and EPPS16~\cite{Eskola:2016oht} PDFs normalized by $\RAA^{i,{\rm iso}}$, given in \Tab{tab:values}.}
\label{fig:raanorm} %TODO show pol0 fit as line with color of points (and width roughly the uncer of the fit err
\end{center} %TODO order EPPS16 and nNNPDF in plot legends
\end{figure}

A more precise comparison of the consistency of the data with the reference calculation is provided in \Fig{fig:raanorm}, which shows the data divided by $\RAA^{i,{\rm ref}}$ as a function of centrality, again with the AP bias of 3\%.
The band around unity illustrates the change of the ratio if instead 2\% or 4\%~ had been assumed for the AP bias, confirming that the data and reference are consistent over full range of centrality.
The ratio can be well described assuming a constant dependence with centrality, with about 5\% suppression relative to the reference, in agreement with the predicted $\RAA^{i,{\rm th}}$, in particular when using the EPPS16~\cite{Eskola:2016oht} nuclear PDFs.
%W+: (0.934062 +/- 0.012591) + (0.000639 +/- 0.000489) * x 
%W-: (0.943325 +/- 0.012217) + (0.001156 +/- 0.000488) * x 
%Z: (0.950231 +/- 0.011920) + (0.000877 +/- 0.000526) * x 
% ----- Fit of double ratio pol0 0 - 100 ----- 
%W+: (0.946578 +/- 0.008168)
%W-: (0.965346 +/- 0.007923)
%Z: (0.965139 +/- 0.007890)  
An alternative way to interpret the ratio shown in the figure is to quantify the nuclear modifications of all three bosons adjusted for isospin effects and biases, where agreement with unity would correspond to ``no nuclear modification''. 
The differing assumptions for an underlying AP bias can then be viewed as a normalization uncertainty increasing with centrality.
%A similar approach was already discussed in \todo{CITE}, where a distinction between two $Q_{\rm AA}$ and \RAA is introduced, where the latter refers to a quantity revised for multiplicity bias.

\begin{figure}[t!]
\begin{center} 
\includegraphics[width=\linewidth]{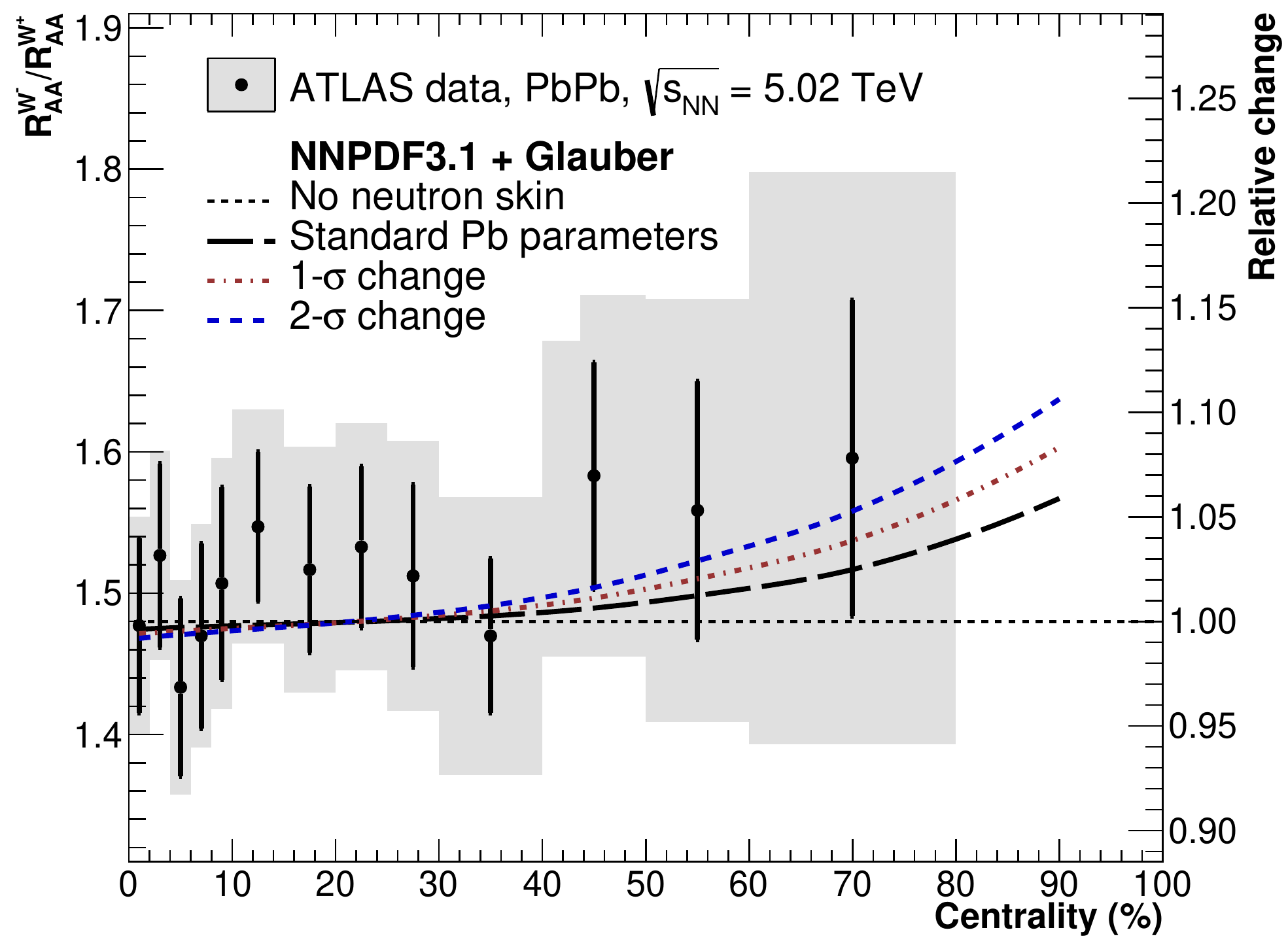}
\caption{$\RAA^{{\rm W}^{-}}/\RAA^{{\rm W}^{+}}$ ratio using the ATLAS data~\cite{Aad:2019sfe} as a function of centrality compared to the same W$^-$/W$^-$ ratio computed with $\RAA^{\rm ref}$.
The $y$-scale on the right, denotes the relative change of the model relative to the baseline~(no neutron skin) given by the respective $\RAA^{\rm iso}$ ratio, shown as a constant line.
In the Glauber model the parameters for the proton and neutron distributions of Pb were modified by one and two standard deviations of their respective uncertainties, as described in the text.
}
\label{fig:raaratio}
\end{center}
\end{figure} % plotskinrat() in macro/playRaa.C
%skin variations 
%5.43455 5.58452 0.149969   cross 7.6b
%5.40852 5.64844 0.239924   cross 7.78b
%5.38266 5.71392 0.331259   cross 7.9b
%5.35699 5.78091 0.423919   cross 8.1b

%%%%%%%%%%%%%%%%%%%%%%%%%%%%%%%%%%%%%%%%
\section{Influence of neutron skin}
\label{sec:neutronskin}
%%%%%%%%%%%%%%%%%%%%%%%%%%%%%%%%%%%%%%%%
To investigate the centrality dependence of the isospin, we study the $\RAA^{{\rm W}^{-}}/\RAA^{{\rm W}^{+}}$ ratio using the ATLAS data~\cite{Aad:2019sfe} as a function of centrality, similarly to an earlier study using an optical Glauber model~\cite{Paukkunen:2015bwa}.
%The sensitivity of $W$ boson production to the presence of a neutron skin was also shown in \Refe{Paukkunen:2015bwa} using an optical Glauber model.  Furthermore, the author suggests that in the future, such measurements could be used to benchmark different centrality definitions at the LHC.
In this ratio, the overall normalization as well as the effects from the possible biases do not play a role, since they fully cancel.
The systematic uncertainties of the data, which are also expected to cancel to some degree, in particular in peripheral collisions, were taken as independent.
In \Fig{fig:raaratio}, the data are compared to the same ratio computed with $\RAA^{\rm ref}$, where essentially everything cancels except the dependence on the isospin resulting from the description of the Pb nucleus from protons and neutrons in the Glauber model.
In {\sc TGlauberMC}~\cite{Loizides:2017ack}, the standard parameters for the radius $r$ and the diffusivity $a$ of the two-parameter fermi distribution of Pb, are $r_{\rm p}=6.68\pm0.02$~fm, $a_{\rm p}=0.447\pm0.01$~fm for the protons, and $r_{\rm n}=6.69\pm0.03$~fm, $a_{\rm n}=0.560\pm0.03$~fm for the neutrons.
With this default parametrization of the Pb nucleus the computed $\RAA^{{\rm W}^{-}}/\RAA^{{\rm W}^{+}}$ ratio exhibits a rather moderate increase of up to 5\%, consistent with the data.
To study the sensitivity of the the isospin effect, the parameters for the proton and neutron distributions of Pb were modified by one and two standard deviations of their respective uncertainties in opposite direction, \ie\ the proton parameters were reduced and the neutron parameters were enhanced, effectively probing different neutron skin thicknesses.
In this way, protons get pushed more to the inside and neutrons more to the outside of the Pb nucleus, which is reflected in the changing difference of the neutron and proton root-mean-square radius $\Delta R$ from 0.15 to 0.24 and 0.33~fm, respectively, consistent with $\Delta R=0.283\pm0.071$~fm~\cite{Adhikari:2021phr} derived from recent measurements of parity-violation in electron scattering.
The investigated variations lead to a relative change of up to 10\% in the expected $\RAA^{{\rm W}^{-}}/\RAA^{{\rm W}^{+}}$ ratio for peripheral collisions compared to the absence of a neutron skin, in good agreement with the data within the uncertainties.
%We note that the investigated changes lead to a smaller mean square charge radius than measured~\cite{Jones:2014aoa}.
Currently, the large uncertainties of the data do not allow to discriminate the Pb nucleus parameters.
However, with factor 10 or more increase of integrated luminosity in Run-3/4 at LHC one certainly will be able to make a more definitive statement.
This is of particular importance as the investigated changes in the Pb parameters also lead to an increase of the \PbPb\ cross section by about 2--4\%~(since neutrons were pushed outwards slightly increasing the overall Pb area).
The visible cross section of the forward plastic scintillating arrays of ALICE at about 50\% centrality was determined~\cite{ALICE-PUBLIC-2021-001}, with about two percent precision to $3.9$b, naively extrapolated leading to $\sigmaPbPb=7.8$b consistent with a few percent increase~(albeit with an uncertainty of similar size) relative to the expected cross section.
In other words, the source of the AP bias may originate from a smaller neutron skin thickness implemented in the Glauber model than recently extracted from the electron parity experiments.
Since all collaborations at the LHC use the same parametrization of the Pb nucleus in the Glauber calculations, it is not clear why the new CMS data~\cite{Sirunyan:2021vjz} do not exhibit a similar problem if the underlying source of the problem stems from the skin thickness.
It is imaginable~(but not possible to quantitatively investigate within the scope of this paper) that the AP bias gets enhanced when relying on the anchor point at 80\%~(ATLAS) instead of 90\%~(CMS); making the ATLAS measurements more sensitive to the correction for missing events than the CMS measurement. 

%%%%%%%%%%%%%%%%%%%%%%%%%%%%%%%%%%%%%%%%
\section{Summary}
\label{sec:summary}
%%%%%%%%%%%%%%%%%%%%%%%%%%%%%%%%%%%%%%%%
Data on the nuclear modification of W and Z bosons measured by the ATLAS collaboration in \PbPb\ collisions at $\snn=\SI{5.02}{TeV}$ show an enhancement in peripheral collisions for all three bosons that was recently explained by arguing that the nucleon--nucleon cross section may be shadowed in nucleus--nucleus collision.
This interpretation has significant consequences for the understanding of heavy-ion data, in particular in the context of the Glauber model, since the the ratio of shadowed over standard inelastic cross section in \pPb\ and \PbPb\ collisions changes as function of collision energy~(\Fig{fig:shcross}), and so one no longer could use the measured $\sigmapp$ cross section as input to the Glauber model.
Instead, we provide an alternative explanation of the data by assuming that there is a mild centrality bias of about 3\% present in the measurement, \ie\ in the determination of the anchor point.
Such an anchor point bias would cancel the multiplicity bias present for hard probes, and effectively result in a slightly rising bias with centrality on the measured boson yield~(\Fig{fig:bias}).
We construct a reference model for the data in absence of nuclear effects by computing the isospin dependence of $\RAA$ using NLO calculations and the Glauber model~(\Eq{eq:model}) together with the bias factor.
The data of the W$^+$, W$^-$ or Z boson \RAA\ measured by the ATLAS collaboration are in agreement with our reference model~(\Fig{fig:raa}); in particular the rising trend in the data can be well explained with our reference model.
Dividing the data by the reference model we extract a rather centrality-independent suppression relative to the unmodified baseline of about 5\%, consistent with NLO calculations using nuclear PDFs, in particular with the EPPS16 PDFs~(\Fig{fig:raanorm}).
The centrality dependence of the $\RAA^{{\rm W}^{-}}/\RAA^{{\rm W}^{+}}$ ratio potentially hints at the relevance of a larger skin thickness of the Pb nucleus than presently included in the Glauber model, although the present precision of the data does not allow for a firm conclusion~(\Fig{fig:raaratio}).
%In other words, the source of the anchor bias may originate from a smaller neutron skin thickness implemented in the Glauber model than recently extracted from the electron parity experiments.
Higher precision data from Run-3/4 and the LHC will be needed to investigate this possibility; these data may also be used as experimental proxy for the nuclear overlap function. 
Furthermore, in order to allow a centrality-determination independent measurement the boson cross section in essentially zero-bias \PbPb\ collisions should be measured. %to allow a direct comparison of cross sections via $\frac{1}{A^2}\frac{{\rm d}\sigma^i_{\rm AA }/{\rm d}p_{\rm{T}}}{{\rm d}\sigma^i_{\rm pp }/{\rm d}p_{\rm{T}}}$. %, which requires to measure the \PbPb\ luminosity using a Van-der-Meer scan.

%%%%%%%%%%%%%%%%%%%%%%%%%%%%%%%%%%%%%%%%
\begin{acknowledgments}
The authors acknowledge financial support by the U.S. Department of Energy, Office of Science, Office of Nuclear Physics, under contract number DE-AC05-00OR22725. One author was furthermore financially supported by the Federal Ministry of Education and Research (BMBF) in the ErUM Framework under contract number 05P19PMCA1.
\end{acknowledgments}
%%%%%%%%%%%%%%%%%%%%%%%%%%%%%%%%%%%%%%%%
\ifextra
\newpage
\section{Appendix}

%\begin{figure}
%\begin{center}
%\includegraphics[width=0.48\textwidth]{./figures/RAA_RatioToCT14.pdf}
%\includegraphics[width=0.48\textwidth]{./figures/RAA_RatioToCT14_zoom.pdf}
%\caption{\todo{Check if we want to keep this plot, maybe change scale of y axis.}}
%\end{center}
%\end{figure}

\begin{figure}
\begin{center}
\includegraphics[width=\textwidth]{./figures/tableATLAS.png}
\caption{Table from ATLAS paper.}
\end{center}
\end{figure}

\begin{table}[t!]
\footnotesize
\caption{Current most recent table (state 14.04.2021). The cross section is calculated via $N_{\text{Bosons,accepted}}\cdot \sigma_{\text{gen.,pythia}}/ N_{\text{events}}$. No correction for the branching ratio is applied. $N_{\text{Bosons,accepted}}$ only includes bosons from decay to $e$ or $\mu$}
\begin{tabular}{cc|SSS}
\toprule
PDF & coll. sys. & {$\sigma_{W^{+}}$ (\si{\pico\barn})}&  {$\sigma_{W^{-}}$ (\si{\pico\barn})} &  {$\sigma_{Z}$ (\si{\pico\barn})} \\
\midrule
 %\parbox[t]{2mm}{\multirow{3}{*}{\rotatebox[origin=c]{90}{\tiny CT14 NLO}}} & pp & \num{1375.504 \pm 5.749} & \num{1029.297 \pm 4.973} & \num{300.939 \pm 1.980} \\
%& pn & \num{1208.497 \pm 5.341} & \num{1215.295 \pm 5.356} & \num{304.815 \pm 1.961} \\
%& nn & \num{1050.165 \pm 4.718} & \num{1377.696 \pm 5.404} & \num{306.851 \pm 1.942} \\
%\midrule
%\parbox[t]{2mm}{\multirow{3}{*}{\rotatebox[origin=c]{90}{\tiny HERAPDF2.0}}}& pp & \num{1409.331 \pm 5.612} & \num{1105.208 \pm 4.970} & \num{313.592 \pm 2.079} \\
%& pn & \num{1270.484 \pm 5.340} & \num{1262.405 \pm 5.323} & \num{317.388 \pm 2.055} \\
%& nn & \num{1106.454 \pm 4.969} & \num{1411.646 \pm 5.612} & \num{323.452 \pm 2.045} \\
\bottomrule
%MCFM NNLO (CT14 NNLO) & pp & \num{2157.83 \pm 2.65} & \num{1556.90\pm1.69} &\num{354.37\pm0.26} \\
MCFM NLO (CT14 NLO) & pp& \num{2154.51\pm0.15} & \num{1348.68\pm0.49} &\num{344.91\pm0.17}\\
MCFM NLO (CT14 NLO) & nn & \num{1527.00 \pm 0.89} & \num{1789.87\pm0.67} &\num{352.64\pm0.16}\\

MCFM NLO (NNPDF31 NLO) & pp& \num{2232.81\pm1.1} & \num{1382.37\pm0.49} &\num{357.72\pm0.17}\\
MCFM NLO (NNPDF31 NLO) & pn& \num{1888.74\pm0.89} & \num{1614.28\pm0.59} &\num{361.93\pm0.17}\\
MCFM NLO (NNPDF31 NLO) & nn& \num{1553.90\pm0.64} & \num{1855.28\pm0.71} &\num{364.48\pm0.17}\\
MCFM NLO (nNNPDF20 A=1) & pp & \num{2213.37\pm0.71} & \num{1342.81\pm0.58} & \num{352.87\pm0.17}\\
%MCFM NLO (CT14 NNLO) & pp& \num{2227.03 \pm 1.08} & \num{1393.20\pm0.50} &\num{354.34\pm0.17}\\
%MCFM LO (CT14 LO) & pp & \num{1843.95 \pm 1.36} & \num{1102.62\pm0.68} &\num{263.37\pm0.11}\\
%MCFM LO (CT14 NLO) & pp & \num{2101.67 \pm 1.52} & \num{1292.07\pm0.66} &\num{305.91\pm0.12}\\
\midrule
MCFM NLO (nNNPDF20) & PbPb & \num{1624.93\pm0.71} & \num{1501.71\pm0.58} & \num{332.83\pm0.17}\\
(values given per nucleon per beam!!) &&&\\
ratio to CT14 NLO pp & PbPb/pp & 0.754 & 1.134 &  0.964  \\
ratio to NNPDF31 pp & PbPb/pp & 0.728 & 1.086 &  0.930  \\
ratio to nNNPDF20 A=1 pp & PbPb/pp & 0.734 & 1.118 &  0.943  \\
 \midrule
 MCFM NLO (EPPS16) & PbPb & \num{1622.20\pm0.71} & \num{1494.74\pm0.58} & \num{330.51\pm0.16} \\
 ratio to MCFM NLO (CT14NLO) & PbPb/pp & 0.753 & 1.108 & 0.958 \\
 \midrule
 MCFM NLO (nCTEQ15) & PbPb & \num{1533.50\pm0.72} & \num{1415.07\pm0.54} & \num{316.36\pm0.15} \\
  MCFM NLO (nCTEQ15) & bound p & \num{1841.87\pm0.88} & \num{1174.21\pm0.48} & \num{311.56\pm0.15} \\
    MCFM NLO (nCTEQ15) & bound n & \num{1336.80\pm0.60} & \num{1575.04\pm0.48} & \num{318.654\pm0.15} \\
 \end{tabular}
\end{table}

\begin{figure}
\begin{center}
\includegraphics[width=8cm]{./figures/cNeutronParams.pdf}
\caption{Phase space for 2PF values for neutrons deduced from $\Delta R=0.282\pm0.071$~fm~\cite{Adhikari:2021phr}.}
\end{center}
\end{figure}

\fi
%%%%%%%%%%%%%%%%%%%%%%%%%%%%%%%%%%%%%%%%
\bibliographystyle{utphys}
\bibliography{biblio}
\end{document}